\documentclass[aps,prb,twocolumn]{revtex4}
\usepackage{graphicx} 
\usepackage{bm}
\usepackage{dcolumn}
\usepackage{amsmath}

\begin{document}
%
\title{Effective  pairing interaction in the two-dimensional Hubbard model \\ 
within a spin rotationally invariant approach}%
%
\author{V. A. Apinyan and T. K. Kope\'{c}}
\affiliation{ Institute for Low Temperature and Structure Research, Polish Academy of Sciences\\
PO. Box 1410, 50-950 Wroclaw 2, Poland \\}
\date{\today}%
\begin{abstract}
We implement the rotationally-invariant formulation of the two-dimensional Hubbard model,
with nearest-neighbors hopping  $t$, which  allows for the analytical study  of the system in the low-energy limit.   Both U(1) and SU(2) gauge transformations are used to factorize the charge and spin contributions to the original electron operator in terms of the 
corresponding  gauge fields.  The  Hubbard Coulomb energy $U$ term is  then expressed 
in terms of quantum phase variables conjugate to the local
charge and variable spin-quantization axis, 
providing a useful representation of strongly correlated systems.
It is shown that these gauge fields play a similar role as phonons in the BCS theory: they
act as the ``glue" for fermion pairing. By tracing out gauge degrees of freedom, the form of paired states is established and  the  strength of the pairing potential is determined. It is found that  the attractive pairing potential in the effective low-energy fermionic action is  non-zero
in a rather narrow range of $U/t$.

\end{abstract}
\maketitle
%
\section{Introduction}
%
Superconductivity represents a remarkable phenomenon where quantum coherence
effects appear at macroscopic scale.\cite{bcs}
The quantum-mechanical  phase of the electrons  gains rigidity and as a result
 the properties of the quantum wave show up at the macro-macroscopic level. 
Thus, the superconducting properties are the manifestations of the spontaneous
breakdown of one of the fundamental symmetries of matter, namely, the U(1)
gauge symmetry. The discovery of the cuprate
superconductors \cite{htsc} has sparked a widespread
interest in physics which goes beyond the traditional
Fermi-liquid framework usually employed for understanding
the effect of interactions in metals.
The question of whether the pairing interaction in the
cuprate superconductors should be characterized as arising
from a ''pairing glue'' has recently been raised.\cite{and}
While there is a growing
consensus that superconductivity in the high-$T_c$ cuprates
arises from strong short-range Coulomb interactions between
electrons rather than the traditional electron-phonon
interaction, the precise nature of the pairing interaction
remains controversial. In this context
the Hubbard model is considered as  essential physical system
for treating superconductivity in the strongly correlated electron systems and
has been intensively studied with a variety of
methods such as quantum Monte-Carlo \cite{hirsch, imada}
(QMC), exact diagonalization,\cite{fano,moreo}
path-integral renormalization-group,\cite{kashima}
functional renormalization-group,\cite{metzner} and various quantum
cluster methods.\cite{cluster} 
As a principal
model describing the electronic correlation in the system, the 
Hubbard model has been used in many works to study the pairing 
instabilities which as usual are given by the second-order effective 
interaction with respect to the Coulomb interaction. 
In this context the structure of the pairing interaction, 
 the two-dimensional (2D) Hubbard model, has been recently analyzed in 
\cite{Scalapino1,Scalapino2,Scalapino4}, where the dynamical cluster Monte Carlo
 approximation is applied to two-dimensional Hubbard model 
with nearest-neighbors hopping and on site Coulomb interaction.
The Monte Carlo simulations have been also employed
to study the phase separation and pairing in the doped two-dimensional 
Hubbard model.\cite{Scalapino3} 
However, the question whether the Hubbard model even supports superconductivity
without additional interactions remains a subject
of controversy. Different mean-field theories suggest
conflicting ground-state order parameters and correlations,
while finite-size QMC simulations for the doped 2D Hubbard
model in the intermediate coupling regime of correlation energy $U$
support in general the idea of a spin-fluctuation-driven interaction mediating
$d$-wave superconductivity. However, the fermion sign problem, 
limits these calculations to temperatures too high to
study a possible transition. These simulations are also
restricted to relatively small system sizes so that the 
off-diagonal long-range order has not
been ascertained.
For theoretical  understanding of the mechanism of superconductivity in cuprates,
the knowledge of bosons mediating the pairing  is of pivotal importance.
Here, the underlying
attraction force appears very puzzling since it is hard to reconcile
the microscopic attractive interaction with the completely
repulsive bare electron-electron forces.
This issue is closely related to the construction of the 
low-energy effective theory for the electronic system. 
A powerful tool for the quantitative investigation of microscopic models is
provided by the study of effective  theories: if one is able to single
out the most relevant low-energy configurations, an effective theory
can be extracted from the microscopic
lattice Hamiltonian. 
This procedure is often implemented via the projective
transformation, which results in removing of high-energy degrees
of freedom and replacing them with kinematical constraints 
as exemplified, e.g., by the $t$-$J$ model.\cite{zhang} 
In such approaches, the high-energy scale associated
with the charge gap is argued to be irrelevant, hence the
focus exclusively on the spin sector to characterize the Mott
insulator. However, the charge-transfer nature of the cuprates
plays an essential role in the doped systems,\cite{za} so that with
discarding charge degrees of freedom an important part of
the physics may be lost.
In the same spirit a detour
from the strict projection program was recently proposed in a
form of the ``gossamer'' superconductor,\cite{la} recognizing the
role of the  double-occupancy charge configurations.
However, the most interesting and relevant situation of strongly correlated systems,
where magnetic as well as charge degrees of freedom interact, was until now investigated to a
much lesser extent since it requires the treatment of the  Hubbard Hamiltonian
without imposing any restrictions on the correlation energy $U$.

In the present paper we study the emergence of the attractive  pairing interaction in the 
two-dimensional Hubbard model by resorting to the analytical method that is deeply  rooted
in the inherent spin-rotational  and  gauge-charge symmetries of the model.
To keep the spin-rotationally invariance, we 
write the action of the system using other bosonic and fermionic 
variables which are introduced with appropriate U(1) and SU(2) transformations.
We construct  a SU(2) spin-rotational and charge U(1) invariant theory using the electron operator 
factorization.\cite{Kopec,Zaleski}
Furthermore, 
we derive the low-energy  fermionic action 
that rests on  the SU(2)-invariant character of the Hamiltonian and 
a consistent scheme of coherent states
within a functional-integral formulation.
We show that U(1) and SU(2)  gauge fields (the collective high energy
modes in the SC system) take over the task which was carried
out by phonons in BCS superconductors and  play the role of the ``glue"
that is responsible for the formation of the electron pairs.
In this sense the present work  charts a route from the microscopic Hubbard
model on the square lattice to an effective lower energy action 
that exhibits pairing potential.
The paper is organized as follows. Section II introduces
the model  and rotational invariant formulation. Section III describes charge and spin
gauge transformations of fermions, which results in the phase-angular representation of
strongly correlated electrons .
 Section IV is devoted to the evaluation of pairing interaction, while
Sec. V discusses the  effective fermionic action. We conclude
with Sec. VI. Appendixes A, B and C contain miscellaneous results that pertain to the
technical aspects of the work.
%
\section{Hubbard model in the rotating reference frame}
%
The basic physics of strongly correlated electronic systems is the
 competition between the two tendencies of the electron to spread out
 as a wave and to localize as a particle, combined with magnetism. 
That is, the interplay of the spin and the charge degree of freedom is the central issue. 
These features are encoded in by the Hubbard Hamiltonian - the simplest
 yet nontrivial model
for strongly correlated electrons.
The relevance of this model for superconducting cuprates  originates 
from the observation that the one–-band Hubbard model tries to mimic the presence
of the charge-transfer gap of cuprates  by means of an effective value of the
Coulomb repulsion.
Thus, our starting point is the purely fermionic Hubbard Hamiltonian in the second-quantized form
\begin{eqnarray}
{\cal H}=
 -t\sum_{\langle {\bf r}{\bf r}'\rangle,\alpha}
 [c^{\dagger }_{{\alpha}}({\bf r})
c_{\alpha }({\bf r}')+{\rm H.c.}]
+
\sum_{\bf r}Un_{\uparrow} ({\bf r}) n_{\downarrow}({\bf r}).
\label{mainham}
\end{eqnarray}
Here, $\langle {\bf r},{\bf r}'\rangle$  runs
over the nearest-neighbor (n.n.) sites, $t$  is the  hopping amplitude, $U$ stands for
the Coulomb repulsion,
while the operator $c_{\alpha }^\dagger({\bf r})$
creates an electron with spin $\alpha=\uparrow(\equiv 1),\downarrow(\equiv 2)$ at the square lattice site ${\bf r}$.
Furthermore,
${n}({{\bf r}})= n_{\uparrow} ({\bf r})+n_{\downarrow}({\bf r})$
is the  number operator, where  ${n}_{\alpha}({{\bf r}})= c^\dagger_{\alpha}({\bf r})
c_{\alpha}({\bf r})$. Usually, working in the grand canonical ensemble  a term
$-\mu\sum_{\bf r}{n}({\bf r})$ is added to ${\cal H}$ in Eq.(\ref{mainham})
with $\mu$  being  the chemical potential .
We treat the problem of interacting fermions at finite
temperature in the standard path-integral formalism\cite{negele}
using Grassmann variables for Fermi fields,
$c_\alpha({\bf r}\tau)$ 
depending on the ``imaginary time" $0\le\tau\le \beta\equiv 1/k_BT$
(with $T$ being the temperature)
that satisfy the anti-periodic condition
$c_{\alpha}({\bf r}\tau)=-c_{\alpha}({\bf r}\tau+\beta)$,
to write the path integral for the statistical sum
${\cal Z}=\int\left[{\cal D}\bar{c}  {\cal D}{c}
\right]e^{-{\cal S}[\bar{c},c]}$
with the fermionic action 
\begin{eqnarray}
{\cal S}[\bar{c},c]={\cal S}_B[\bar{c},c]+\int_0^\beta d\tau{\cal H}[\bar{c},c],
 \end{eqnarray}
that contains the fermionic  Berry term
\begin{eqnarray}
{\cal S}_B[\bar{c},c]=\sum_{{\bf r}\alpha }\int_0^\beta d\tau
 \bar{c}_{\alpha }({\bf r}\tau)\partial_\tau{c}_{\alpha }({\bf r}\tau)
\label{cberry}.
 \end{eqnarray}
For the problem under study  it is  crucial to construct a covariant formulation of the theory,
which  naturally  preserves  the  spin-rotational symmetry present in the Hubbard Hamiltonian.
For this purpose, the density-density product 
in Eq.(\ref{mainham}) we write, following Ref. \onlinecite{schulz}, in
a spin-rotational invariant way
\begin{equation}
{\cal H}_U=U\sum_{{\bf r} }\left\{\frac{1}{4}{n }^2({{\bf r}}\tau)
-\left[{\bf \Omega} ({\bf r}\tau)\cdot{\bf S} ({\bf r}\tau)\right]^2\right\},
\label{huu}
\end{equation}
where
 $S^a({\bf r}\tau)=\frac{1}{2}\sum_{\alpha\alpha'}c^\dagger_{\alpha}({\bf r}\tau)
\hat{\sigma}_{\alpha\alpha'}^a c_{\alpha'}({\bf r}\tau)$
denotes the vector spin operator ($a=x,y,z$) with $\hat{\sigma}^a$ being
the Pauli matrices. The unit vector 
 ${\bf \Omega} ({\bf r}\tau)=[\sin\vartheta({\bf r}\tau)\cos\varphi({\bf r}\tau),
\sin\vartheta({\bf r}\tau)\sin\varphi({\bf r}\tau),
\cos\vartheta({\bf r}\tau)]$
written in terms of polar angles labels
varying in space-time  spin-quantization axis. 
 Thus, the Hubbard 
 Hamiltonian should not change its form under a rotation of the spin-quantization axis. This is not apparent in the standard form of the interaction in Eq.(\ref{mainham}).
The spin-rotation invariance is made explicit
by performing the angular integration over  ${\bf\Omega}({\bf r}\tau)$
at each site and time. By decoupling spin- and charge-density terms 
in Eq.(\ref{huu}) using auxiliary fields
$\varrho({\bf r}\tau)$ and $iV({\bf r}\tau)$, respectively,
we write down the partition function in the form,
\begin{eqnarray}
{\cal Z}&=&\int[{\cal D}{\bf \Omega}]\int[{\cal D} V{\cal D}{\varrho}
 ]\int\left[
  {\cal D}\bar{c}{\cal D}c
\right]\times
\nonumber\\
&\times&
e^{-{\cal S}\left[{\bf \Omega},V,{\varrho},\bar{c},c\right]},
\label{zfun}
\end{eqnarray}
where$[{\cal D}{\bf \Omega}]\equiv
\prod_{{\bf r}\tau_k}
\frac{\sin\vartheta({\bf r}\tau_k)d\vartheta ({\bf r}\tau_k)d\varphi ({\bf r}\tau_k)}{4\pi}$
is the spin-angular integration measure.
The  effective action reads as
\begin{eqnarray}
{\cal S}\left[{\bf \Omega},V,{\varrho},\bar{c},c\right]&=&
\sum_{ {\bf r} }\int_0^\beta
 d\tau
\left[\frac{{\varrho}^2 ({\bf r}\tau)}{U}+\frac{V^2 ({\bf r}\tau)}{U}\right.
\nonumber\\
&+&\left.iV ({\bf r}\tau)n ({\bf r}\tau)
+2{\varrho} ({\bf r}\tau){\bf \Omega} ({\bf r}\tau)\cdot {\bf S} ({\bf r}\tau)\right]
\nonumber\\
&+&{\cal S}_B[\bar{c},c]+\int_0^\beta d\tau{\cal H}_t[\bar{c},c].
\label{sa}
\end{eqnarray}
We would like to stress that the fermionic fields in Eq. (\ref{sa}) are
the physical ones and not due to an enlargement of the Hilbert space like in a slave-boson
treatment of the $t$-$J$ model.\cite{zhang} As we see in Secs.IIIA-IIIC,
 the gauge fields will arise here by relating an SU(2) rotation 
in spin space and a vector on the two sphere ($S_2$). 
%
\section{Charge and spin gauge transformations of fermions}
%
The interaction term of the Hubbard Hamiltonian is quartic in fermion operators.
This is   a nonlinear problem which is not solvable except in some
very special cases, such as one-dimensional systems. Thus, a standard
approach is the mean-field approximation, sometimes called Hartree-Fock
(HF)approximation, in which the quartic term is factorized in terms 
of a fermion bilinear times an auxiliary field, which is usually treated classically.
However, HF theory will not work for a Hubbard model in which $U$
is the largest energy in the problem.
One has to isolate strongly fluctuating modes generated by the Hubbard term  according to the
charge-U(1) and spin-SU(2) symmetries.
%
\subsection{U(1) charge transformation}
%
We swich mow
from the particle-number representation to the conjugate
phase representation of the electronic degrees of freedom.
To this aim the second-quantized
Hamiltonian of the model is translated to the phase
representation with the help of the topologically constrained
path-integral formalism. 
To this end  we write the fluctuating ``imaginary chemical 
potential" $iV ({\bf r}\tau)$ as a sum of
a static $V_{0 }({\bf r})$ and periodic function
$V({\bf r}\tau)=V_0({\bf r})+\tilde{V}({\bf r}\tau)$
using Fourier series,
\begin{eqnarray}
\tilde{V}({\bf r}\tau)=\frac{1}{\beta}	\sum_{n=1}^\infty
[\tilde{V}({\bf r}\omega_n)e^{i\omega_n\tau}+\text{c.c.}]
\label{decomp}
\end{eqnarray}
with $\omega_n=2\pi n/\beta$ ($n=0,\pm1,\pm2$)
being the (Bose) Matsubara frequencies.
Now, we introduce the U(1) {\it phase } field ${\phi} ({\bf r}\tau)$
via the Faraday-type relation,
\begin{equation}
\dot{\phi} ({\bf r}\tau)\equiv\frac{\partial\phi ({\bf r}\tau)}
{\partial\tau}=\tilde{V} ({\bf r}\tau).
\label{jos}
\end{equation}
Since the homotopy group $\pi_1$[U(1)] forms a set of integers,
discrete configurations of $\phi({\bf r}\tau)$ matter, for which
$\phi({\bf r}\beta)-\phi({\bf r}0)=2\pi m({\bf r})$, where $m({\bf r})=0,\pm 1,\pm 2,\dots$
Thus the decomposition  of the charge field
$V({\bf r}\tau)$   conforms with the  basic $m=0$ topological sector since
$\int_0^\beta\dot{\phi} ({\bf r}\tau)=\int_0^\beta \tilde{V} ({\bf r}\tau)\equiv 0$.
Furthermore, by performing the local gauge transformation to the {\it new} fermionic
variables $f_{\alpha}({\bf r}\tau)$, 
\begin{eqnarray}
\left[\begin{array}{c}
c_{\alpha }({\bf r}\tau)\\
\bar{c}_{\alpha }({\bf r}\tau)
\end{array}\right]=
\left[\begin{array}{cc}
z({\bf r}\tau)&0\\
0& \bar{z}({\bf r}\tau)
\end{array}\right]
\left[\begin{array}{c}
f_{\alpha }({\bf r}\tau)\\
\bar{f}_{\alpha }({\bf r}\tau)
\end{array}\right]
\label{sing1}
\end{eqnarray}
where the unimodular parameter  $|z({\bf r}\tau)|^2=1$ satisfies $z({\bf r}\tau)=e^{i\phi ({\bf r}\tau)}$,
we remove the imaginary term $i\int_0^\beta d\tau\tilde{V}({\bf r}\tau)n({\bf r}\tau)$
for all the Fourier modes
of the $V ({\bf r}\tau)$ field, except for  the zero frequency.
%
\subsection{SU(2) spin transformation}
%
In the above description, we focused on the charge degree of freedom 
of the electron. However, the electron has one more degree of freedom being the spin. The spin dominates the magnetic properties. 
The subsequent SU(2) transformation from $f_{\alpha}({\bf r}\tau)$ to  $h_{\alpha}({\bf r}\tau)$
variables,
\begin{eqnarray}
\left[\begin{array}{c}
f_{1 }({\bf r}\tau)\\
{f}_{2}({\bf r}\tau)
\end{array}\right]=
\left[
\begin{array}{cc}
\zeta_{ 1}({\bf r}\tau) & -\bar{\zeta}_{2}({\bf r}\tau) \\
\zeta_{2}({\bf r}\tau) & \bar{\zeta}_{1}({\bf r}\tau)
\end{array}
\right]\left[\begin{array}{c}
h_{1 }({\bf r}\tau)\\
{h}_{2}({\bf r}\tau)
\end{array}\right]
\label{sing2}
\end{eqnarray}
with the constraint
$|\zeta_{1}({\bf r}\tau)|^2 +|\zeta_{2}({\bf r}\tau)|^2=1$ takes away the
rotational dependence on ${\bf \Omega}({\bf r}\tau)$ in the spin sector.
This is done
by means  of the Hopf map,
\begin{equation}
{\bf R}({\bf r}\tau) \hat{\sigma}^z{\bf R}^\dagger({\bf r}\tau)
 =\hat{{\bm\sigma}}\cdot{\bf \Omega}({\bf r}\tau)
\end{equation}
where
\begin{equation}
{\bf R}({\bf r}\tau)=\left[
\begin{array}{cc}
\zeta_{ 1}({\bf r}\tau) & -\bar{\zeta}_{2}({\bf r}\tau) \\
\zeta_{2}({\bf r}\tau) & \bar{\zeta}_{1}({\bf r}\tau)
\end{array}
\right]
\end{equation}
that is based on the enlargement from two sphere $S_2$ to the three sphere $S_3\sim {\rm SU}(2)$.
The unimodular constraint
can be resolved by using the parametrization
\begin{eqnarray}
\zeta_{1}({\bf r}\tau)& = & e^{-{i}/{2}[\varphi({\bf r}\tau)+\chi({\bf r}\tau)]}
\cos\left[\frac{\vartheta({\bf r}\tau)}{2}\right]
\nonumber\\
\zeta_{2}({\bf r}\tau)&=&e^{{i}/{2}[\varphi({\bf r}\tau)-\chi({\bf r}\tau)]}
\sin\left[\frac{\vartheta({\bf r}\tau)}{2}\right]
\label{cp1}
\end{eqnarray}
with the  Euler
angular variables $\varphi({\bf r}\tau),\vartheta({\bf r}\tau)$ and $\chi({\bf r}\tau)$, respectively.
Here, the  extra variable $\chi({\bf r}\tau)$   represents the U(1) gauge freedom  of the theory
as a consequence of $S_2\to S_3$ mapping. One can summarize Eqs. (\ref{sing1}) and (\ref{sing2})
by the single joint gauge transformation exhibiting electron operator factorization
\begin{eqnarray}
c_{\alpha }({\bf r}\tau)=\sum_{\alpha'}z({\bf r}\tau)
{ R}_{\alpha\alpha'} ({\bf r}\tau)h_{\alpha'}({\bf r}\tau),
\label{decomp2}
\end{eqnarray}
where ${\bf R}({\bf r}\tau) =e^{-i\hat{\sigma}_z\varphi({\bf r}\tau)/2}e^{-i\hat{\sigma}_y\vartheta({\bf r}\tau)/2}
e^{-i\hat{\sigma}_z\chi({\bf r}\tau)/2}$
is  a unitary matrix which
rotates the spin-quantization axis at site ${\bf r}$ and time $\tau$.
Equation(\ref{decomp2}) reflects the composite nature of the interacting electron formed from
bosonic spinorial and charge degrees of freedom given by  ${ R}_{\alpha\alpha'} ({\bf r}\tau)$
and $z({\bf r}\tau)$, respectively, as well as  remaining fermionic core part $h_{\alpha}({\bf r}\tau)$.
%
\subsection{Fermionic sector}
%
Anticipating that spatial and temporal fluctuations of the fields $V_0({\bf r})$
and $\varrho({\bf r }\tau)$ will be energetically penalized,
since they are gaped and decouple from the angular and phase variables.
Therefore, in order to make further progress 
we  subject the functional  in Eq.(\ref{sa}) to a  saddle-point analysis with respect to non-fluctuating (static) fields
and variables that fluctuations cost energy of the order of  $U$. 
The expectation value of the static (zero-frequency) part of
the fluctuating potential $V_0(r)$ in the charge sector  we calculate
by the saddle-point method to gives
\begin{equation}
V_0(r)=i\left( \mu-\frac{U}{2}n_h \right)\equiv i\bar{\mu}
\end{equation}
where $\bar{\mu}$ is the chemical potential
with a Hartree shift originating from the saddle-point value of the static
 variable $V_0({\bf r})$ with $n_h=n_{h\uparrow}+n_{h\downarrow}$ and $n_{h\alpha}=
\langle\bar{h}_{\alpha }({\bf r}\tau)h_{\alpha }({\bf r}\tau)\rangle$.
Similarly in the magnetic sector, 
a saddle-point evaluation of $\rho({\bf r})$  reproduces the conventional
 Hartree–Fock equations for a commensurate antiferromagnet
\begin{eqnarray}
\rho({\bf r}\tau)=
(-1)^{\bf r}\Delta_c
\label{spaff}
\end{eqnarray}
where $\Delta_c=U\langle S^z({\bf r}\tau) \rangle$ sets the magnitude for the 
Mott-charge gap  $\Delta_c\sim U/2$ for $U/t\gg 1$.
The staggerization factor
$(-1)^{\bf r}$  breaks the translation invariance by one site which remains by two sites.
The term  is readily handled by going to the reduced Brillouin zone.\cite{Zaleski}	
Note that the notion antifferomagnetic here does not mean
an actual  long-range ordering - for this the angular spin-quantization variables 
governed by the rotational symmetry have to be ordered as well.
To summarize, the fermionic sector is governed by the effective Hamiltonian
\begin{eqnarray}
 &&{\cal H}_{\bf \Omega,\phi} = 
	\sum_{{ \bf r}}\Delta_c (-1)^{\bf r} [\bar{h}_{{\uparrow} }({\bf r}\tau)h_{\uparrow  }({\bf r}\tau)-
\bar{h}_{{\downarrow} }({\bf r}\tau)h_{\downarrow  }({\bf r}\tau)]
\nonumber\\
&&-t\sum_{\langle {\bf r}{\bf r}'\rangle,\alpha\gamma} \bar{z}({\bf r}\tau)z({\bf r}'\tau)
 \left[{\bf R}^\dagger ({\bf r}\tau){\bf R}_{ }({\bf r'}\tau)\right]_{\alpha\gamma}
\bar{h}_{{\alpha} }({\bf r}\tau)h_{\gamma }({\bf r}'\tau)
\nonumber\\
&&-\bar{\mu}\sum_{{ \bf r}\alpha}
\bar{h}_{\alpha }({\bf r}\tau)
 h_{\alpha }({\bf r}\tau),
\label{explicit}
\end{eqnarray}
The chief merit of the gauge transformation
in Eq.(\ref{decomp2}) is that we have managed to cast the Hubbard
problem into a system of  $h$ fermions
submerged in the bath of strongly fluctuating U(1) and SU(2) gauge potentials (minimally coupled to 
fermions via hopping term) which mediate the interactions.
%
\section{Pairing interaction}
%
It is well known that the crucial point of BCS theory is the existence of 
an attractive interaction among electrons, where  that phonons play the 
role of the ``glue" that is responsible for the formation of  Cooper pairs.
 Here, by integration out the bosonic scalar filed that represents phonons 
in the fermionic Hamiltonian, an effective attractive
potential emerges.
Now we show that U(1) {\it and} SU(2)
emergent gauge fields (the collective high-energy modes in the Hubbard system)  take
over the task which was carried out by phonons in BCS superconductors.
In a way similar to phonons these  gauge fields  couple to the fermion  density-type term 
via the   amplitude $t$, see Eq.(\ref{explicit}),
\begin{equation}
-t\sum_{\langle {\bf r}{\bf r}'\rangle,\alpha\gamma} \bar{z}({\bf r}\tau)z({\bf r}'\tau)
 \left[{\bf R}^\dagger ({\bf r}\tau){\bf R}_{ }({\bf r'}\tau)\right]_{\alpha\gamma}
\bar{h}_{{\alpha} }({\bf r}\tau)h_{\gamma }({\bf r}'\tau)
\end{equation}
Thus, in order to obtain  an effective interaction among fermions  we
have to integrate out all the bosonic modes  given by $\bar{z}({\bf r}\tau),z({\bf r}'\tau)$
and ${\bf R}^\dagger ({\bf r}\tau),{\bf R}_{ }({\bf r'}\tau) $ fields. A major difference with 
respect to the BCS theory is that the variables to be integrated out are of tensorial nature
since SU(2) modes carry spin index.
To explicitly evaluate the effective interaction between fermions by
tracing out the gauge degrees of freedom, we resort to the cumulant expansion
To this end we write the partition function as ${\cal Z}=\int[{\cal D}{\bar h}{\cal D}h]e^{-{\cal S}[\bar{h},h]}$,
where the effective fermionic action is 
\begin{eqnarray}
S_{eff}[\bar{h},h]=-\ln\int\left[{\cal{D}}\phi{\cal{D}}
{\bf{\Omega}}\right]e^{-S\left[{\bf{\Omega}},\phi,\bar{h},h\right]}
\label{cex}
\end{eqnarray}
The expression Eq.(\ref{cex}) generates a cumulant series when expanded with respect
to the hopping variable $t$. The relevant second-order term that contains the quartic fermionic
term becomes
\begin{widetext}
\begin{eqnarray}
&&S^{(2)}\left[\bar{h},h\right]=-{t^{2}}\sum_{
\left\langle{\bf{r}}_{1}{\bf{r}}'_{1}\right\rangle}\sum_{
\left\langle{\bf{r}}_{2}{\bf{r}}'_{2}\right\rangle}\int^{\beta}_{0}d\tau{d\tau'}
\left\langle \frac{}{} \bar{z}({\bf{r}}_{1}\tau)z({\bf{r}}'_{1}
\tau)\bar{z}({\bf{r}}_{2}\tau')z({\bf{r}}'_{2}\tau')\frac{}{}\right\rangle_{\text{U(1)}}
\nonumber\\
&&\times\sum_{\alpha\alpha'}\sum_{\gamma\gamma'}\left\langle\frac{}{} 
\left[{\bf{R}}^{\dag}({\bf{r}}_{1}\tau){\bf{R}}({\bf{r}}'_{1}\tau)
\right]_{\alpha\alpha'}\left[{\bf{R}}^{\dag}({\bf{r}}_{2}\tau'){\bf{R}}({\bf{r}}'_{2}\tau')
\right]_{\gamma\gamma'}\right\rangle_{\text{SU(2)}}
{\bar{h}_{\alpha}({\bf{r}}_{1}\tau)h_{\alpha'}({\bf{r}}'_{1}\tau)}
\bar{h}_{\gamma}({\bf{r}}_{2}\tau')h_{\gamma'}({\bf{r}}'_{2}\tau')
\label{secondordercumulant}
\end{eqnarray} 
\end{widetext}
where 
\begin{equation}
\left\langle \dots\right\rangle_{\text{U(1)}}=\frac{\int\left[{\cal{D}}\phi\right]\dots
e^{-S\left[\phi\right]}}{\int\left[{\cal{D}}\phi\right]
e^{-S\left[\phi\right]}}
\end{equation}
 is the averaging over U(1) phase field while 
\begin{equation}
\left\langle\dots\right\rangle_{\text{SU(2)}}=\frac{\int\left[{\cal{D}}{\bf{\Omega}}\right]
e^{-S\left[{\bf{\Omega}}\right]}}{\int\left[{\cal{D}}{\bf{\Omega}}\right]
e^{-S\left[{\bf{\Omega}}\right]}}
\end{equation}
 is the averaging over spin-angular variables. 
To proceed with the evaluation of the effective fermion-fermion interaction one has to develop
procedures for effectuating both averages which involves calculation of the effective actions
$S\left[\phi\right]$ and $S\left[{\bf{\Omega}}\right]$ in charge and spin sectors, respectively.	
%
\subsection{U(1) average}
%
The averaging in the charge sector is performed with the use of the U(1) phase action (see Appendix A).
\begin{eqnarray}
S[\phi]=\sum_{\bf r}\int_0^\beta d\tau\left[\frac{\dot{\phi}^2({\bf r}\tau)}{U} 
+\frac{2\mu}{iU}\dot{\phi}({\bf r}\tau) \right]
\label{sphi}
\end{eqnarray}
that contains both  the kinetic  and Berry terms of the U(1) phase field in the charge sector.
For the U(1) average in Eq.(\ref{secondordercumulant}) we get
\begin{eqnarray}
\left\langle \frac{}{} \bar{z}({\bf{r}}_{1}\tau)z({\bf{r}}'_{1}\tau)
\bar{z}({\bf{r}}_{2}\tau')z({\bf{r}}'_{2}\tau')\frac{}{}\right
\rangle_{\text{U(1)}}\approx \left(\delta_{{\bf{r}}_{1}{\bf{r}}'_{1}}
\delta_{{\bf{r}}_{2}{\bf{r}}'_{2}}\right.
\nonumber\\
\left.+\delta_{{\bf{r}}_{1}{\bf{r}}'_{2}}\delta_{{\bf{r}}'_{1}{\bf{r}}_{2}}
\right)e^{-\frac{U}{2}\left[|\tau-\tau'|-{\left(\tau-\tau'\right)^{2}}/{\beta}\right]}\label{pppp}. 
\end{eqnarray} 
Specializing to  the low-temperature limit
\begin{eqnarray}
\lim_{\tau\rightarrow 0}\int^{\beta}_{0}d\tau'e^{-|\tau-\tau'|{U}/{2}}
=\lim_{\tau\rightarrow 0}\left[\frac{2}{U}-\frac{2e^{-\beta{U}}}{2}\right]
= \frac{2}{U},
\end{eqnarray}
we obtain the result for the U(1) phase average.
\subsection{ SU(2) average}
%
\subsubsection{Spin-angular action}
The calculation of the SU(2) average  is done with help of  the
 effective action that involves the spin-directional degrees of 
freedom ${\bf \Omega}$, which important fluctuations correspond
to rotations. This can be done by integrating out fermions
 ${\cal Z}=\int\left[{\cal D}{\bf \Omega}\right]e^{-{\cal S}[{\bf \Omega}]}$
 where 
\begin{eqnarray}
{\cal S}[{\bf \Omega}]=-\ln\int[{\cal D}\phi{\cal D}\bar{h}{\cal D}h]
e^{-{\cal S}[\varphi,\phi,\vartheta,\bar{h},h]}
\end{eqnarray}
generates the  low-energy action in the
form 
${\cal S}[{\bf \Omega}]=\mathcal{S}_{0}\left[\mathbf{\Omega}\right]
+{\cal S}_{B}[{\bf \Omega}]+{\cal S}_{J}[{\bf \Omega}]$.
The interaction term with the spin stiffness becomes
\begin{equation}
{\cal S}_{J}[{\bf \Omega}] =  \frac{J\left(\Delta\right)}{4}
\sum_{\langle{\bf r}{\bf r}'\rangle}\int_{0}^{\beta}d\tau
{\bf \Omega}({\bf r}\tau)\cdot{\bf \Omega}({\bf r'}\tau),
\label{eq:SJ}
\end{equation}
with the entiferromagnetic (AF) exchange coefficient 
\begin{eqnarray}
 &  & J(\Delta_{c})=\frac{4t^{2}}{U}(n_{\uparrow}
-n_{\downarrow})^{2}\equiv\frac{4t^{2}}{U}\left(\frac{2\Delta_{c}}{U}\right)^{2}.\label{afex}
\end{eqnarray}
 From Eq. (\ref{afex}) it is evident that for $U\to\infty$ one
has $J(\Delta_{c})\sim\frac{4t^{2}}{U}$ since $\frac{2\Delta_{c}}{U}\to1$
in this limit.
Thus, in the strong-coupling limit, the half-filled Hubbard model
 maps onto the quantum Heisenberg model. In this limit the 
fermions are bound into localized spins. There is no motion 
of fermions, since they are suppressed  by the gap for charge fluctuations.
 In general the AF-exchange parameter persists as long
as the charge gap $\Delta_{c}$ exists. However, $J(\Delta_{c})$
diminishes rapidly in the $U/t\to0$ weak-coupling limit.
Because the gauge field is the phase factor arising in the 
inner product of quantum-mechanical states - the so-called 
connection in mathematical language - it is intimately 
related to the Berry phase term $ {\cal S}_{B}[{\bf \Omega}]$ in the effective action.
If we work in Dirac ``north pole\char`\" gauge $\chi({\bf r}\tau)=-\varphi({\bf r}\tau)$
one recovers the familiar form 
\begin{equation}
{\cal S}_{B}[{\bf \Omega}]=\frac{\theta}{i}\sum_{{\bf r}}
\int_{0}^{\beta}d\tau\dot{\varphi}({\bf r}\tau)[1-\cos\vartheta({\bf r}\tau)].
\label{sberry}
\end{equation}
Here, the integral on the right-hand side of  Eq. (\ref{sberry}) has a
simple geometrical interpretation as it is equal to a solid angle
swept by a unit vector ${\bf \Omega}(\vartheta,\varphi)$ during its
motion.\cite{auer} The extra phase factor coming from the Berry phase requires
some little extra care, since it will induce quantum-mechanical phase
interference between configurations. In regard to the nonperturbative
effects, we realized the presence of an additional parameter with
the topological angle or so-called theta term 
\begin{equation}
\theta=\frac{\Delta_{c}}{U}\label{topoltheta}
\end{equation}
that is related to the Mott gap. In the large-$U$ limit, one has
$\Delta_{c}\to U/2$, so that $\theta\to\frac{1}{2}$ relevant for
the half-integer spin.
The kinetic-energy term in the spin sector becomes
\begin{eqnarray}
{\cal S}_0[{\bf \Omega}]&=&
\sum_{\bf r}\int_0^\beta d\tau\left\{\right.
\frac{1}{4{\cal E}_s}
\left[{ \dot{\vartheta}^2({\bf r}\tau)+\dot{\varphi}^2({\bf r}\tau)+\dot{\chi}^2({\bf r}\tau)}
\right.
\nonumber\\
&+&
\left. {2}\dot{\varphi}({\bf r}\tau)
\dot{\chi}({\bf r}\tau)\cos{\vartheta}({\bf r}\tau)
\right]
\left.\right\},
\label{phiaction2}
\end{eqnarray}
where $\mathcal{E}_{s}=1/\left(2\chi_{T}\right)$ and 
\begin{equation}
\chi_{T}= \left\{
\begin{array}{cc}
\displaystyle\frac{1}{8J} & \,\,\, t\ll U\\
\displaystyle\frac{1}{2\pi}\frac{1}{t}\sqrt{\frac{t}{U}} & \,\,\, t\gg U
\end{array}\right.
\end{equation}
is the transverse spin suusceptibility.
%
\subsubsection{CP$^{1}$ representation}
%
It the $CP^{1}$ representation, the spin-quantization axis can be
conveniently written as
\begin{equation}
\mathbf{\Omega}\left(\mathbf{r}\tau\right)=\sum_{\alpha\alpha'}
\bar{\zeta}_{\alpha}\left(\mathbf{r}\tau\right)
\bm{\sigma}_{\alpha\alpha'}\zeta_{\alpha'}\left(\mathbf{r}\tau\right).
\end{equation}
As a consequence, all the terms in the spin action can be expressed
as functions of unimodular $\zeta_{\alpha}\left(\mathbf{r}\tau\right)$
variables instead of angular variables, which are more complicated
to be handled.
Finally, the action assumes the form
\begin{eqnarray}
 &  & \mathcal{S}\left[\bar{\bm{\zeta}},\bm{\zeta}\right]
=\sum_{\mathbf{r}}\int_{0}^{\beta}d\tau\left\{ 2\chi_{T}
\dot{\bm{\zeta}}\left(\mathbf{r}\tau\right)\cdot
\dot{\bm{\zeta}}\left(\mathbf{r}\tau\right)\right.	
\nonumber \\
 &  & -\left.\theta\left(-1\right)^{\mathbf{r}}
\left[\bar{\bm{\zeta}}\left(\mathbf{r}\tau\right)
\cdot\dot{\bm{\zeta}}\left(\mathbf{r}\tau\right)
-\dot{\bar{\bm{\zeta}}}\left(\mathbf{r}\tau\right)
\cdot\bm{\zeta}\left(\mathbf{r}\tau\right)\right]
\right\} \nonumber \\
 &  & -J\sum_{\left\langle \mathbf{r}\mathbf{r}'
\right\rangle }\int_{0}^{\beta}d\tau
\bar{\mathcal{A}}\left(\mathbf{r}\tau\mathbf{r}'\tau\right)
\mathcal{A}\left(\mathbf{r}\tau\mathbf{r}'\tau\right)
\label{eq:Szetabzeta}
\end{eqnarray}
with the bond operators
\begin{eqnarray}
 &  & \bar{\mathcal{A}}\left(\mathbf{r}\tau\mathbf{r}'\tau\right)
\mathcal{A}\left(\mathbf{r}\tau\mathbf{r}'\tau\right)
=-\frac{1}{4}\mathbf{\Omega}\left(\mathbf{r}\tau\right)
\cdot\mathbf{\Omega}\left(\mathbf{r}'\tau\right)+\frac{1}{4}\nonumber \\
 &  & \mathcal{A}\left(\mathbf{r}\tau\mathbf{r}'
\tau\right)=\frac{\zeta_{\uparrow}\left(\mathbf{r}\tau\right)
\zeta_{\downarrow}\left(\mathbf{r}'\tau\right)
-\zeta_{\downarrow}\mathbf{\left(\mathbf{r}\tau\right)}
\zeta_{\uparrow}\left(\mathbf{r}'\tau\right)}{\sqrt{2}}
\label{aaa}.
\end{eqnarray}
%
\subsubsection{Canonical transformation of CP$^{1}$ variables}
%
In order to achieve a consistent representation of the underlying
antiferromagnetic structure, it is unavoidable to explicitly split
the degrees of freedom according to their location on sublattice A
or B. Since the lattice is bipartite, allowing one to make the unitary
transformation 
\begin{eqnarray}
\zeta_{\uparrow}({\bf r}\tau) & \to & -\zeta_{\downarrow}({\bf r}\tau)\nonumber \\
\zeta_{\downarrow}({\bf r}\tau) & \to & \zeta_{\uparrow}({\bf r}\tau)
\label{cantra}
\end{eqnarray}
for sites on one sublattice, so that the antiferromagnetic bond operator
becomes
\begin{eqnarray}
{\cal A}({\bf r}\tau{\bf r'}\tau) & \to & {\cal A}'({\bf r}\tau{\bf r'}\tau
)=\sum_{\alpha=1}^{2}\frac{\zeta_{\alpha}({\bf r}\tau)\zeta_{\alpha}({\bf r}'\tau)}{\sqrt{2}}.
\end{eqnarray}
This canonical transformation preserves the unimodular constraint  of the $CP^{1}$ fields.
Biquadratic (four-variable) terms in the Lagrangian cannot be readily
integrated in the path integral. Introducing a complex variable for
each bond that depends on ``imaginary time\char`\" $Q({\bf r}\tau{\bf r'}\tau)$,
we decouple the four-variable terms $\bar{{\cal A}}'({\bf r}\tau{\bf r'}\tau){\cal A}'({\bf r}\tau{\bf r'}\tau)$
using the formula
\begin{eqnarray}
e^{{\cal S}_{J}[\bar{\bm{\zeta}},\bm{\zeta}]} & = & \int[{\cal D}^{2}Q]
e^{-\sum\limits _{\langle{\bf r}{\bf r}'\rangle}\int
\limits _{0}^{\beta}d\tau\left(\frac{2}{J}|Q|^{2}+Q\bar{\bm{\zeta}}
\cdot\bar{\bm{\zeta}}+\bar{Q}\bm{\zeta}\cdot\bm{\zeta}\right)}\nonumber \\
{\cal D}^{2}Q & = & \prod_{\langle{\bf r}{\bf r'}\rangle\tau}d^{2}Q({\bf r}\tau{\bf r'}\tau),
\end{eqnarray}
where $d^{2}Q=d{\rm Re}Qd{\rm Im}Q$. In a similar manner by introducing
a local real field $A\left(\mathbf{r}\tau\right)$, we can decouple the
second term in the right-hand side. in Eq. (\ref{eq:Szetabzeta}). To handle
the unimodularity condition, one introduces a Lagrange multiplier $\lambda_{\zeta}(\tau)$.
Then with the help of the Dirac-delta functional,
\begin{equation}
\delta\left(\sum_{\mathbf{r}}|\bm{\zeta}({\bf r}\tau)|^{2}-N\right)
=\int\left[\frac{{\cal D}\lambda_{\zeta}}{2\pi i}\right]
e^{\int\limits _{0}^{\beta}d\tau\lambda_{\zeta}\left(\sum_{\mathbf{r}}|{\bm{\zeta}}|^{2}-N\right)}\end{equation}
where the variables $\zeta_{\uparrow}({\bf r}\tau)$ and $\zeta_{\downarrow}({\bf r}\tau)$
are now unconstrained bosonic fields. Thus, the local constraints
are reintroduced into the theory through the dynamical fluctuations
of the auxiliary $\lambda_\zeta$ field, so that the statistical sum becomes
\begin{eqnarray}
{\cal Z} & = & \int[{\cal D}^{2}Q{\cal D}^{2}\bm{\zeta}{\cal D}\lambda_{\zeta}]\times\nonumber \\
 & \times & e^{-\sum\limits _{\langle{\bf r}{\bf r}'\rangle}\int\limits _{0}^{\beta}d\tau
\left(\frac{2|Q|^{2}}{J}-\lambda_{\zeta}\delta_{{\bf r}{\bf r}'}+{\cal H}_{Q}[\bar{\bm{\zeta}},{\bm{\zeta}}]\right)},
\end{eqnarray}
where
 \begin{eqnarray}
{\cal H}_{Q}[\bar{\bm{\zeta}},{\bm{\zeta}}] & = & 
\sum_{\langle{\bf r}{\bf r}'\rangle}\int_{0}^{\beta}d\tau
\left[\lambda_{\zeta}{\bar{\bm{\zeta}}}\cdot{\bm{\zeta}}\delta_{\mathbf{r}\mathbf{r}'}\right.
\nonumber \\
 & + & \left.Q{\bar{\bm{\zeta}}}\cdot\bar{\bm{\zeta}}+\bar{Q}{\bm{\zeta}}\cdot{\bm{\zeta}}\right].
\end{eqnarray}
 Furthermore, by evaluating saddle-point values of the $Q$, $a$,
and $\lambda_{\zeta}$ fields 
\begin{eqnarray}
Q_{{\rm sp}}({\bf r}\tau{\bf r}'\tau) & = & -\frac{J}{2}\langle\bar{\bm{\zeta}}({\bf r}\tau)
\cdot\bar{\bm{\zeta}}({\bf r}'\tau)\rangle,\nonumber \\
1 & = & \langle\bar{\bm{\zeta}}({\bf r}\tau)\cdot\bm{\zeta}({\bf r}\tau)\rangle
\end{eqnarray}
and by assuming the uniform solutions $Q_{{\rm sp}}({\bf r}\tau{\bf r}'\tau)\equiv Q$,
$a_{\text{sp}}\left(\mathbf{r}\tau\mathbf{r}'\tau\right)\equiv a$, and 
$\lambda_{\zeta \text{sp}}\left(\tau\right)\equiv\lambda_{\zeta}$,
we obtain for the Hamiltonian in the spin-bosonic sector\begin{equation}
\mathcal{H}\left[\bar{\bm{\zeta}},\bm{\zeta}\right]
=\frac{1}{2\beta N}
\sum_{\mathbf{k}n\sigma}
\bar{\Lambda}_{\zeta\sigma}\left(\mathbf{k}\omega_{n}\right)
G_{\zeta0\mathbf{k}}^{-1}\left(\omega_{n}\right)
\Lambda_{\zeta\sigma}\left(\mathbf{k}\omega_{n}\right)\label{eq:SpinSector_finalaction}\end{equation}
with
\begin{equation}
\Lambda_{\zeta\sigma}\left(\mathbf{k}\omega_{n}\right)=\left[\begin{array}{c}
\zeta_{\sigma}\left(\mathbf{k},\omega_{n}\right)\\
\bar{\zeta}_{\sigma}\left(-\mathbf{k},-\omega_{n}\right)\\
\zeta_{\sigma}\left(\mathbf{k}-\bm{\pi},\omega_{n}\right)\\
\bar{\zeta}_{\sigma}\left(-\mathbf{k}+\bm{\pi},-\omega_{n}\right)\end{array}\right]
\label{prop}
\end{equation}
and
\begin{equation}
G_{\zeta0\mathbf{k}}^{-1}\left(\omega_{n}\right)=\left[\begin{array}{cccc}
\frac{\omega_{n}^{2}}{\mathcal{E}_{s}}+\lambda_{\zeta} & 2Q\xi_{\mathbf{k}} & -2i\theta\omega_{n} & 0\\
2Q\xi_{\mathbf{k}} & \frac{\omega_{n}^{2}}{\mathcal{E}_{s}}+\lambda_{\zeta} & 0 & 2i\theta\omega_{n}\\
-2i\theta\omega_{n} & 0 & \frac{\omega_{n}^{2}}{\mathcal{E}_{s}}+\lambda_{\zeta} & -2Q\xi_{\mathbf{k}}\\
0 & 2i\theta\omega_{n} & -2Q\xi_{\mathbf{k}} & \frac{\omega_{n}^{2}}{\mathcal{E}_{s}}+\lambda_{\zeta}
\end{array}\right].
\label{ppp}
\end{equation}
Decoupling of the bond operators in the kinetic term of the
spin action in Eq. (\ref{eq:Szetabzeta}) leads to additional field
$Q$, which value is determined from the equation,
\begin{eqnarray}
Q=\frac{J(\Delta)}{zN}\sum_{\mathbf{k}}\xi_{\mathbf{k}}W_{\mathbf{k}}
\label{qcalc}
\end{eqnarray}
while the constraint parameter $\lambda_\zeta$ is the solution of the equation
\begin{eqnarray}
1=\frac{J(\Delta)}{N}\sum_{\mathbf{k}}W_{\mathbf{k}}
\end{eqnarray}
with
\begin{eqnarray}
 &  & W_{\mathbf{k}}=
 \frac{\coth\left[\beta E_{s\mathbf{k}}^{-}\left(\omega_{\mathbf{k}}^{-}\right)
\right]+\coth\left[\beta E_{s\mathbf{k}}^{+}\left(\omega_{\mathbf{k}}^{-}\right)\right]}
{4\sqrt{\theta^{2}+\frac{\omega_{\mathbf{k}}^{-}}{\mathcal{E}_{s}}}}\nonumber \\
 &  & +\frac{\coth\left[\beta E_{s\mathbf{k}}^{-}\left(\omega_{\mathbf{k}}^{+}\right)\right]
+\coth\left[\beta E_{s\mathbf{k}}^{+}\left(\omega_{\mathbf{k}}^{+}\right)\right]}
{4\sqrt{\theta^{2}+\frac{\omega_{\mathbf{k}}^{+}}{\mathcal{E}_{s}}}} ,
\end{eqnarray}
where:
\begin{equation}
E_{s\mathbf{k}}^{\pm}\left(\omega_{\mathbf{k}}^{\pm}\right)=
\frac{\mathcal{E}_{s}}{2}\left(\sqrt{\theta^{2}
+\frac{\omega_{\mathbf{k}}^{\pm}}{\mathcal{E}_{s}}}
\pm\theta\right)\end{equation}
and\begin{equation}
\omega_{\mathbf{k}}^{\pm}=\lambda_{\zeta}\pm2Q\xi_{\mathbf{k}}
\end{equation}
with $\xi_{\mathbf{k}}=\frac{1}{2}[\cos(k_x)+\cos(k_y)]$ as the two-dimensional lattice
structure factor.
%
\subsection{Fermionic action}
%
We start the calculation of the effective fermionic action with the
first-order term with respect to the hopping element $t$
\begin{eqnarray}
&&S^{(1)}_{t}=-t\sum_{\left\langle \textbf{r}\textbf{r}'
\right\rangle}\left\langle \bar{z}(\textbf{r}\tau)z(\textbf{r}'\tau)\right\rangle_{\text{U(1)}}
\nonumber\\
&&\times \left\langle 
\left[R^{\dag}(\textbf{r}\tau)R(\textbf{r}'\tau)\right]_{\alpha\beta}
\right\rangle_{\text{SU(2)}}\bar{h}_{\alpha}(\textbf{r}\tau)h_{\beta}(\textbf{r}'\tau)
\end{eqnarray} 
The evaluation of the average with rotational matrices gives
\begin{widetext}
\begin{eqnarray}
&&\left\langle \sum_{\alpha\beta}[R^{\dag}(\textbf{r}
\tau)R(\textbf{r}'\tau)]_{\alpha\beta}\right\rangle_{\text{SU(2)}} \bar{h}_{\alpha}
(\textbf{r}\tau)h_{\beta}(\textbf{r}'\tau)=\sum_{\alpha}
\left\langle  \zeta_{\alpha}(\textbf{r}\tau){\zeta}_{\alpha}
(\textbf{r}'\tau)\right\rangle_{\text{SU(2)}} \bar{h}_{\downarrow}(\textbf{r}\tau)h_{\uparrow}(\textbf{r}'\tau)
\nonumber\\
&&-\sum_{\alpha}\left\langle  \bar{\zeta}_{\alpha}
(\textbf{r}\tau)\bar{\zeta}_{\alpha}(\textbf{r}'\tau)
\right\rangle_{\text{SU(2)}} \bar{h}_{\uparrow}(\textbf{r}\tau)
h_{\downarrow}(\textbf{r}'\tau)=\sum_{\alpha}
\left\langle 
 \zeta_\alpha(\textbf{r}\tau){\zeta}_\alpha(\textbf{r}'\tau)
\right\rangle[\bar{h}_{\downarrow}(\textbf{r}\tau)h_{\uparrow}
(\textbf{r}'\tau)-\bar{h}_{\uparrow}(\textbf{r}\tau)h_{\downarrow}(\textbf{r}'\tau)]
\end{eqnarray}
\end{widetext}
The first-order action is then in the form
\begin{equation}
S^{(1)}_{t}=-\tilde{t}\sum_{\left\langle \textbf{r}\textbf{r}'\right\rangle}\int_0^\beta
[\bar{h}_{\downarrow}(\textbf{r}\tau)h_{\uparrow}(\textbf{r}'\tau)
-\bar{h}_{\uparrow}(\textbf{r}\tau)h_{\downarrow}(\textbf{r}'\tau)],
\label{1st}
\end{equation}
where $\tilde{t}=tg_c({\bf d})g_s({\bf d})$,  is the renormalized hopping, with   $g_c({\bf d})
=\left\langle \bar{z}(\textbf{r}\tau)z(\textbf{r}'\tau)\right\rangle_{\text{U(1)}}$ 
and $g_s({\bf d})=\sum_\alpha\left\langle \zeta_{\alpha}({\bf{r}}\tau)\zeta_{\alpha}({\bf{r}}'\tau)
\right\rangle_{\text{SU(2)}}$ being the Gutzwiller-type charge and spin renormalization factors.
%
\subsection{Second-order contribution to the fermionic action}
%
The calculation of the second-order contribution to the effective fermionic 
action in Eq.(\ref{secondordercumulant}) is more involved since the SU(2)
averages contain tensorial quantities of the form
\begin{eqnarray}
&&{\bf{M}}_{\alpha\alpha',\gamma\gamma'}({\bf{r}}
\tau,{\bf{r}}'\tau|{\bf{r}}'\tau,{\bf{r}}\tau)=
\nonumber\\
&&=\left\langle\frac{}{} \left[{\bf{R}}^{\dag}
({\bf{r}}_{1}\tau){\bf{R}}({\bf{r}}'_{1}\tau)\right]_{\alpha\alpha'}
\left[{\bf{R}}^{\dag}({\bf{r}}_{2}\tau'){\bf{R}}({\bf{r}}'_{2}\tau')\right]_{\gamma\gamma'}\right\rangle_{\text{SU(2)}}.
\end{eqnarray}
The sublattice  transformation  of the $CP^{1}$ variables in Eq.(\ref{cantra})
translates  to the transformation of the rotation matrix ${\bf{R}}({\bf{r}}\tau)
\to {\bf{\widetilde R}}({\bf{r}}\tau)$ matrix
\begin{eqnarray}
{\bf{R}}({\bf{r}}\tau)=(i\hat{\bf{\sigma}}_{y})\widetilde{\bf{R}}({\bf{r}}'\tau),
\end{eqnarray} 
where ${\widetilde{\bf{R}}}({\bf{r}}\tau)$ is the transformed form of the rotation matrix
\begin{eqnarray}
\widetilde{\bf{R}}({\bf{r}}\tau)=\left[
\begin{array}{cccc}
-\zeta_{2}({\bf{r}}
\tau) & -\bar{\zeta}_{1}({\bf{r}}\tau)
\\ \zeta_{1}({\bf{r}}\tau) & -\bar{\zeta}_{2}({\bf{r}}\tau)
\end{array}
\right].
\end{eqnarray}
It is convenient to define the following bond operator constructed from  the $CP^{1}$ fields:
\begin{equation}
{\cal{F}}({\bf{r}}\tau{\bf{r}}'\tau)=\frac{\bar{\zeta}_{1}({\bf{r}}\tau)
{\zeta}_{1}({\bf{r}}'\tau)+\bar{\zeta}_{2}({\bf{r}}\tau){\zeta}_{2}({\bf{r}}'\tau)}{\sqrt{2}}.
\label{fff}
\end{equation}
With the dedinition in Eq.(\ref{fff}) the 
matrix  ${\bf{M}}_{\alpha\alpha',\gamma\gamma'}
({\bf{r}}\tau,{\bf{r}}'\tau|{\bf{r}}'\tau,{\bf{r}}\tau) $ 
will be written in a compact form as
\begin{eqnarray}
&&{\bf{M}}_{\alpha\alpha',\gamma\gamma'}(\textbf{r}\tau ,\textbf{r}'\tau |\textbf{r}'\tau,\textbf{r}\tau)=
\nonumber\\
&=&\left\langle\left[
\begin{array}{ccccrrrr}
{\cal{F}}\bar{{\cal{F}}} & {\cal{F}}\bar{{\cal{A}}} & -{\cal{F}}{\cal{A}} &  {\cal{F}}{\cal{F}}\\
-\bar{{\cal{A}}}\bar{{\cal{F}}} &-\bar{{\cal{A}}}\bar{{\cal{A}}}  &
 \bar{{\cal{A}}}{\cal{A}} & -\bar{{\cal{A}}}{\cal{F}} \\
{\cal{A}}\bar{{\cal{F}}} & {\cal{A}}\bar{{\cal{A}}} & 
-{\cal{A}}{\cal{A}} & {\cal{A}}{\cal{F}} \\
\bar{{\cal{F}}}\bar{{\cal{F}}} & \bar{{\cal{F}}}\bar{{\cal{A}}} &
 -\bar{{\cal{F}}}{\cal{A}} & \bar{{\cal{F}}}{\cal{F}}  
\end{array}
\right]_{\alpha\alpha',\gamma\gamma'}\right\rangle_{\text{SU(2)}},
\end{eqnarray}
where ${\alpha\alpha',\gamma\gamma'}=\{11,12,21,22$\}.
Now, we can rewrite the second-order fermionic action taking into account the  non-vanishing
averages over $CP^{1}$ fields (see Appendix C) to get
\begin{widetext}
\begin{eqnarray}
&&S^{(2)}[\bar{h},h]=-\frac{t^{2}}{{\cal{U}}}\int^{\beta}_{0}d\tau
\sum_{\left\langle {\bf{r}}{\bf{r}}'\right\rangle}M_{11,11}({\bf{r}}\tau,{\bf{r}}'
\tau|{\bf{r}}'\tau,{\bf{r}}\tau)\sum_{\alpha}\bar{h}_{\alpha}({\bf{r}\tau})
h_{\alpha}({\bf{r}}'\tau)\bar{h}_{\alpha}({\bf{r}}'\tau)h_{\alpha}({\bf{r}\tau})
\nonumber\\
&&+M_{11,22}({\bf{r}}\tau,{\bf{r}}'\tau|{\bf{r}}'\tau,{\bf{r}}\tau)
\left[\sum_{\alpha\beta}\bar{h}_{\alpha}({\bf{r}}\tau)h_{\alpha}({\bf{r}}'\tau)
\bar{h}_{\beta}({\bf{r}}'\tau)h_{\beta}({\bf{r}\tau})-\sum_{\alpha}\bar{h}_{\alpha}({\bf{r}\tau})
h_{\alpha}({\bf{r}}'\tau)\bar{h}_{\alpha}({\bf{r}}'\tau)h_{\alpha}({\bf{r}\tau})\right]
\nonumber\\
&&+M_{12,21}({\bf{r}}\tau,{\bf{r}}'\tau|{\bf{r}}'\tau,{\bf{r}}\tau)
\left[\sum_{\alpha\beta}\bar{h}_{\alpha}({\bf{r}\tau})h_{\beta}({\bf{r}}'\tau)
\bar{h}_{\beta}({\bf{r}}'\tau)h_{\alpha}({\bf{r}\tau})-\sum_{\alpha}\bar{h}_{\alpha}({\bf{r}\tau})
h_{\alpha}({\bf{r}}'\tau)\bar{h}_{\alpha}({\bf{r}}'\tau)h_{\alpha}({\bf{r}\tau})\right]
\nonumber\\
&&+M_{12,12}({\bf{r}}\tau,{\bf{r}}'\tau|{\bf{r}}'\tau,{\bf{r}}\tau)
\left[\sum_{\alpha\beta}\bar{h}_{\alpha}({\bf{r}\tau})h_{\beta}({\bf{r}}'\tau)
\bar{h}_{\alpha}({\bf{r}\tau}')h_{\beta}({\bf{r}\tau})
-\sum_{\alpha}\bar{h}_{\alpha}({\bf{r}\tau})h_{\alpha}({\bf{r}}'\tau)\bar{h}_{\alpha}({\bf{r}}'\tau)
h_{\alpha}({\bf{r}\tau})\right]
\label{s2act}
\end{eqnarray}
\end{widetext}
In deriving the above result, we made the observation
that  the  dynamics of spin variables is slower 
as compared to the charge counterparts,
allowing to treat  SU(2) variables as local in time
${\bf R}({\bf r}\tau')={\bf R}_{ }({\bf r}\tau)+(\tau'-\tau)\partial_\tau
{\bf R}({\bf r}\tau) + O[(\tau'-\tau)^2]$ which leads to nonretarded interactions.\cite{retard}
Furthermore, with the help of the operator identities from Appendix C we can  reduce 
Eq.(\ref{s2act}) to a compact form
\begin{widetext}
\begin{eqnarray}
S^{(2)}[\bar{h},h]=\frac{t^{2}}{{{U}}}\int_0^\beta{d\tau}
\sum_{\left\langle {\bf{r}}{\bf{r}}'\right\rangle}\left[\gamma_{1}n({\bf{r}\tau})n({\bf{r}'\tau})+
\gamma_{2}\bar{\cal A}_h'({\bf{r}\tau}{\bf{r}'\tau}){\cal A}_h'({\bf{r}\tau}{\bf{r}'\tau})
+\gamma_{3}{\bf{S}}_h	({\bf{r}\tau})\cdot{\bf{S}}_h({\bf{r}}'\tau)+
\gamma_{4}n({\bf{r\tau}})
\right],
\end{eqnarray}
\end{widetext}
where the interaction coefficients 
\begin{eqnarray}
&&\gamma_{1}=f^{2}({\bf 0})+2g^{2}({\bf 0})+g^{2}({\bf{d}})+4f^{2}({\bf{d}})>0,
\nonumber\\
&&\gamma_{2}=-2[6f^{2}({\bf{d}})+2f^{2}({\bf 0})]<0,
\nonumber\\
&&\gamma_{3}=4[f^{2}({\bf 0})-g^{2}({\bf{d}})],
\nonumber\\
&&\gamma_{4}=2g^{2}({\bf{d}})+2f^{2}({\bf{d}})+4g^{2}({\bf 0})>0,
\label{gamms}
\end{eqnarray}
%
\begin{figure}
\begin{center}
\includegraphics[scale=0.8]{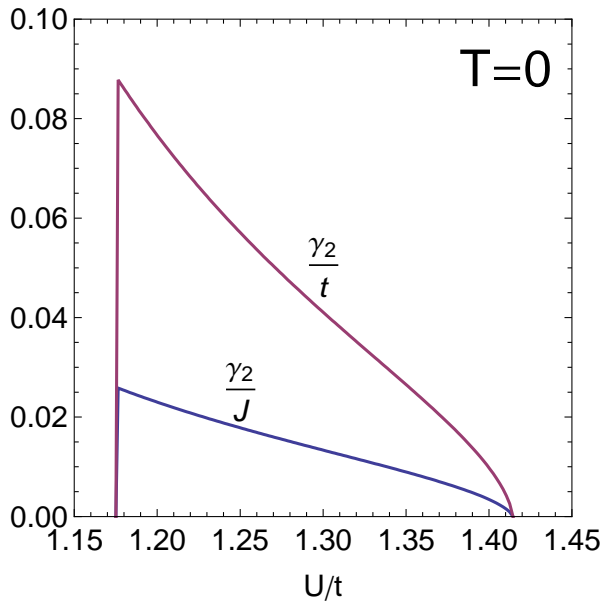}
\caption{(Color online) Pairing interaction $\gamma_2$ normalized to the hopping parameter $t$ (upper curve) and the antiferromagnetic-exchange parameter $J=4t^2/U$ (lower curve)
as a function of the Coulomb interaction $U/t$ calculated at zero temperature  and
half filling $ \bar{\mu}=0$ for the two-dimensional
Hubbard model with nearest-neighbors hopping.}
\end{center}
\end{figure}
%
are given in terms   of the  $CP^{1}$ normal $(g)$   and anomalous $(f)$ correlation functions 
\begin{eqnarray}
&&g({\bf r-r'})=-\left\langle \zeta_\alpha({\bf{r}}\tau)
\bar{\zeta}_{\alpha}({\bf{r}'}\tau)\right\rangle_{\text{SU(2)}},\ 
\nonumber\\
&&f({\bf r-r'})=\left\langle \zeta_{\alpha}({\bf{r}}\tau)\zeta_{\alpha}({\bf{r}}'\tau)
\right\rangle_{\text{SU(2)}}
\label{latticepropagators} 
\end{eqnarray}
which can be readily evaluated using	the propagator  of the $\zeta$ fields in Eq.(\ref{prop}).
%
\section{Hamiltonian with pairing term}
%
From the result in Eq.(\ref{s2act}) we can deduce 
the spin-singlet pairing possibility in the fermionicsector. 
To bring the kinetic-energy term to a standard form, one performs
a rotation of the fermionic variables on one of the sublattices in a manner similar to
the bosonic transformation in Eq.(\ref{cantra}).
\begin{eqnarray}
h_{\uparrow}({\bf{r}}'\tau)\rightarrow -h_{\downarrow}({\bf{r}}'\tau),
\nonumber\\
h_{\downarrow}({\bf{r}}'\tau)\rightarrow h_{\uparrow}({\bf{r}}'\tau).\ \
\end{eqnarray}
As a result the hopping term  assumes the conventional form that is diagonal in the spin indices
\begin{eqnarray}
S^{(1)}_{t}[\bar{h},h]=-\tilde{t}\sum_{\left\langle \bf{r}\bf{r}'\right\rangle, \alpha}\int_0^\beta
\bar{h}_{\alpha}(\textbf{r}\tau)h_{\alpha}(\textbf{r}'\tau),
\end{eqnarray}
while the second-order term is given by
\begin{eqnarray}
S^{(2)}[\bar{h},h]=\sum_{\left\langle {\bf{r}}{\bf{r}}'\right\rangle}\int^{\beta}_{0}d{\tau}
\left[\gamma_1 n({\bf{r}}\tau)n({\bf{r}}'\tau)\right.
\nonumber\\
\left.-\gamma_{2}\bar{\cal A}_h({\bf{r}}\tau{\bf{r}}'\tau){{\cal A}_h}({\bf{r}}\tau{\bf{r}}'\tau)\right],
\label{secondorder}
\end{eqnarray}
where
\begin{eqnarray}
&&{\cal A}_h({\bf{r}}\tau{\bf{r}}'\tau)
=\frac{h_{\uparrow}({\bf{r}}\tau)h_{\downarrow}({\bf{r}}'\tau)
-h_{\downarrow}({\bf{r}}\tau)h_{\uparrow}({\bf{r}}'\tau)}{\sqrt{2}},
\nonumber\\
&&\bar{\cal A}_h({\bf{r}}\tau{\bf{r}}'\tau)
=\frac{\bar{h}_{\downarrow}({\bf{r}}'\tau)\bar{h}_{\uparrow}({\bf{r}}\tau)
-\bar{h}_{\uparrow}({\bf{r}}'\tau)\bar{h}_{\downarrow}({\bf{r}}\tau)}{\sqrt{2}}
\end{eqnarray}
are the bond operators relevant for a singlet  pairing.
The rotational invariance of the right-hand side in
Eq.(\ref{secondorder}) is manifest since
\begin{eqnarray}
&&-\bar{\cal A}_h({\bf r}\tau{\bf r}'\tau){\cal A}_h({\bf r}\tau{\bf r}'\tau)=
\nonumber\\
&&{\bf S}_{h}({\bf r}\tau)\cdot{\bf S}_{h}({\bf r'}\tau )-\frac{1}{4}n_h({\bf r}\tau)n_h({\bf r'}\tau ).
\end{eqnarray}
The coefficients $\gamma_{1}$ and $\gamma_{2}$
are given by  Eq.(\ref{gamms}). By noting that $g({\bf d})=0$ and $f({\bf 0})=0$ one  obtains
\begin{eqnarray}
\gamma_{1}=\frac{4t^{2}}{U}\left[f^2({\bf d})+\frac{1}{2}g^2({\bf 0})\right],
\nonumber\\
\gamma_{2}=\frac{4t^{2}}{U}\left[3f^2({\bf d})\right]=J\left[\frac{3Q^2}{J^2(\Delta)}\right],
\end{eqnarray}
where the $f,g$-correlation functions can be computed with the help of the $CP^{1}$ propagators:
see Eq.(\ref{ppp}).
The effective nonretarded interaction containing $\gamma_2$ in front of the  
$\bar{\cal A}({\bf r}\tau{\bf r}'\tau){\cal A}({\bf r}\tau{\bf r}'\tau)$ term is negative
and therefore constitutes the attractive potential for fermion pairing.
 We can see  that the coefficient $\gamma_2$ is not just given by the 
bare AF exchange $J=4t^2/U$ but is renormalized  downward by the quantity $Q$ that is related to the
antifferomagnetic spin stiffness as delineated in Sec.IVD dealing with  the
SU(2) spin sector. We have calculated $Q$ self-consistently using Eq.(\ref{qcalc}). The result
is plotted in Fig.1. Note that the pairing interaction survives in rather  narrow range of
the Coulomb interaction $1.17<U/t<1.41$. This result suggests that  superconductivity 
in the Hubbard model, if possible, represents a rather delicate balance between
kinetic  energy and Coulomb interaction. In this context, we note that the deleterious 
effects of the Coulomb interaction superconductivity in cuprates  have been largely ignored in the 
literature. Furthermore, since $\gamma_1>0$ in the density-density term in Eq.(\ref{secondorder}),
many sorts of the charge-ordered states can be stabilized, including
{\it e.g.}, charge-density wave states, which in general compete with superconductivity.
This is in contrast to the BCS theory where the only instability
of a Fermi liquid is the Cooper instability: the superconducting order is generic. 
%
\section{CONCLUSIONS and perspectives}
%
The basic physics of strongly correlated electronic systems is
the competition between the two tendencies of the electron to 
spread out as a wave and to localize as a particle
combined with magnetism. That is, the interplay of the spin 
and the charge degree of freedom is the central issue.
While there is a growing consensus that superconductivity 
in the cuprates arises from strong short-range Coulomb interactions
between electrons rather than the traditional
electron-phonon interaction, the precise nature of the
pairing interaction remains controversial.
While the principal focus of the present work is theoretical, the choice of
 model and the approach is motivated 
in the experimentally observed properties of cuprates. 
Therefore, in the present work we hope to shed some light on this controversial 
issue with the purpose to understand better the physical properties 
of most common model for cuprates.	
To this end, we have discussed 
in the present work  the Hubbard model in the spin-rotational
invariant formulation which observes the important symmetries involved.
We presented a field-theoretic description of a microscopic model 
that  reveals an intimate relationship between the spin-SU(2) and  charge-U(1) symmetry and pairings.
We found  that the maximal strength of the effective pairing interaction
parameter is observed in a rather narrow range of $U/t$ with the kinetic energy 
comparable  to  the Coulomb interaction. 
Moreover, the form of the effective fermionic action suggests  
that other competing ordered phases can occur simultaneously, which can quench 
the superconductivity substantially. Therefore, the issue of pairing interaction
 is not settling the question about the long-range superconducting order
in the Hubbard model. As far as modeling of cuprates is concerned there is also a problem
of interplane interaction, entirely omited in the present work, which can 
affect the bulk superconductivity considerably.
In closing we note that the
pairing interaction itself cannot be measured directly: one needs to analyze
key experiments which reveal fingerprints of it.
Thus, the continuing
experimental search for a pairing glue in the cuprates is
important and will play an essential role in determining the
origin of the high-$T_c$ pairing interaction.

\section{ACKNOWLEDGMENTS}
One of us (V.A.A.) acknowledges the financial support 
from the International Max Planck Research School
``Dynamical Processes in Atoms, Molecules, and Solids'',
T.K.K was  supported  by the Ministry of Education and Science
MEN under Grant No. 1PO3B 103 30 in the years 2006-2008.
We are grateful to T. A. Zaleski for numerical evaluation and plotting of the data for
Fig.1.
%
%

%
\section{APPENDIX A:U(1) PHASE AVERAGES}
%
In this section we evaluate the expression for U(1) phase propagator. 
Two point phase-phase propagator for the Bosonic phase
variables is defined as
\begin{eqnarray}
g_{z}({\bf{r}}\tau{\bf{r}}'\tau')
=\left\langle \bar{z}({\bf{r}}\tau)z({\bf{r}}'\tau')\right\rangle.
\end{eqnarray}
The averaging in this definition is over the U(1) phase field and 
\begin{eqnarray}
\left\langle ...\right\rangle=\frac{\int{D\phi}...e^{-S_{0}[\phi]}}{\int{D\phi}e^{-S_{0}[\phi]}}.
\end{eqnarray}
Here the complex variables $z({\bf{r}}\tau)$ 
are defined 
as $z({\bf{r}}\tau)=e^{i\phi({\bf{r}}\tau)}$. 
The variables $\phi({\bf{r}}\tau)$ satisfy the 
following boundary conditions:
\begin{eqnarray}
\phi({\bf r}\beta)-\phi({\bf r}0)=2{\pi}m({{\bf r}}).
\end{eqnarray}
It is very convenient to satisfy the boundary conditions by decomposing
the phase field in terms of a periodic field $\phi({\bf{r}}\tau)$
 and a term linear in $\tau$. We set
\begin{eqnarray}
\phi({\bf r}\tau)=\tilde{\phi}({\bf r}\tau)+\frac{2\pi\tau}{\beta}m({\bf r})
\end{eqnarray}
with 
$\tilde{\phi}(\beta)=\tilde{\phi}(0)$. Summing over the phase field $\varphi$ 
means to integrate all $\phi({\bf{r}}\tau)$ configurations and perform the summation over the integers $n$.
Then we write the phase variables $\phi({\bf{r}}\tau)$ 
in the Fourier-transformed form
\begin{equation}
\tilde{\phi}({\bf{r}}\tau)=\frac{\phi_{0}({\bf{r}})}{\beta}
+\frac{1}{\beta}\sum^{\infty}_{n=1}\left[\phi_{n}({\bf{r}})
e^{i\omega_{n}\tau}+\phi^{\ast}({\bf{r}})e^{-i\omega_{n}\tau}\right].
\end{equation}
The weight of the averaging in the expression of the phase correlator
is given by the following exponential:

\begin{eqnarray}
e^{-S_{0}[\phi]}=e^{-\sum_{{\bf{r}}}\int^{\beta}_{0}d\tau
{\dot{\phi}^{2}({\bf{r}}\tau)}/{U}\dot{\phi}}e^{-S_{0}[m]},
\label{equ1}
\end{eqnarray}
where the action $S_{0}[m]$ is the topological part of the action and is given by
\begin{eqnarray}
S_{0}[\phi ] &=& \frac{2}{\beta{U}}\sum_{{\bf{r}}}\sum_{n}|\phi_{n}({\bf{r}})|^{2},
\nonumber\\
S_{0}[m]&=&\sum_{{\bf{r}}}\left[\frac{4\pi^{2}m^{2}({\bf r})}
{\beta{U}}-\frac{4\pi{i}\mu}{U}m({\bf r})\right].
\end{eqnarray}
Now we evaluate the average. We write first the nontopological part of the action
\begin{widetext}
\begin{eqnarray}
g_{z}({\bf{r}}\tau{\bf{r}}'\tau')&=&\frac{\int[{\cal{D}}\phi]
e^{-i\phi({\bf{r}}\tau)}e^{i\phi({\bf{r}}\tau)}e^{-{2}/{\beta{U}}
\sum_{{\bf{r}}}\sum_{n}|\phi_{n}({\bf{r}}\tau)|^{2}}}
{\int[{\cal{D}}\phi]e^{-{2}/{\beta{U}}\sum_{{\bf{r}}}\sum_{n}|\phi_{n}({\bf{r}})|^{2}}}
\nonumber\\
&&= \delta_{\bf rr'}\frac{\prod^{\infty}_{n=1}\left[\frac{U\beta}
{4\omega^{2}_{n}}\right]e^{-{U}/{2\beta}\sum^{\infty}_{n=1}
{1}/{\omega^{2}_{n}}\left\{\left[\sin(\omega_{n}\tau)
\sin(\omega_{n}\tau')\right]^{2}+\left[\cos(\omega_{n}\tau)
-\cos(\omega_{n}\tau')\right]^{2}\right\}}}
{\prod^{\infty}_{n=1}\left[\frac{U\beta}{4\omega^{2}_{n}}\right]}
\end{eqnarray}
\end{widetext}
By using the identity
\begin{eqnarray}
&&\left[\sin (x)-\sin( y)\right]^{2}+\left[\cos(x)-\cos(y)\right]^{2}
\nonumber\\
&&=2-2[\cos(x)\cos(y)+\sin(x)\sin(y)]
\nonumber\\
&&=2-2\cos(x-y)
\end{eqnarray}
one obtains
\begin{equation}
g_{z}({\bf{r}}\tau{\bf{r}}'\tau')=\delta_{\bf rr'}
 e^{-{U}/{\beta}\sum^{\infty}_{n=1}{1}/{\omega^{2}_{n}}\left\{1-\cos\left[\omega_{n}(\tau-\tau')\right]\right\}}.
\end{equation}
Now, in order to calculate the sum in the exponential we use the following identity:
\begin{eqnarray}
|x|-\frac{x^{2}}{\beta}=\sum^{\infty}_{n=1}\left(\frac{4}{\beta\omega^{2}_{n}}
-\frac{4\cos(\omega_{n}\tau)x}{\beta\omega^{2}_{n}}\right),
\end{eqnarray}
where $-\beta \leq x \leq \beta$. And finally we get

\begin{eqnarray}
g_{z}({\bf{r}}\tau{\bf{r}}'\tau')=\delta_{\bf rr'}
e^{-{U}/{4}\left[|\tau-\tau'|-{(\tau-\tau')^{2}}/{\beta}\right]}
\label{pp}
\end{eqnarray}
Now we are ready to calculate the four-point phase correlator:
\begin{eqnarray}
&&\left\langle \bar{z}({\bf{r}}_{1}\tau)z({\bf{r}}'_{1})
\bar{z}({\bf{r}}_{2}\tau')z({\bf{r}}_{2}\tau')\right\rangle_{\text{U(1)}} = 
\nonumber\\
&&\left\langle \bar{z}({\bf{r}}_{1}\tau)z({\bf{r}}'_{1})
\right\rangle_{\text{U(1)}} \left\langle \bar{z}({\bf{r}}_{2}\tau')
z({\bf{r}}_{2}\tau')\right\rangle_{\text{U(1)}}+
\nonumber\\
&&	 + \left\langle \bar{z}({\bf{r}}_{1}\tau)z({\bf{r}}'_{2}\tau')
\right\rangle_{\text{U(1)}} \left\langle z({\bf{r}}'_{1}\tau)
\bar{z}({\bf{r}}_{2}\tau')\right\rangle_{\text{U(1)}}
\end{eqnarray}
By using the result in Eq. (\ref{pp}) we get  Eq.(\ref{pppp}).
%
%
\section{APPENDIX B:SU(2) AVERAGE}
%
%
\begin{widetext}
The composition formula for the rotational matrices in the angular representation is given by
\begin{eqnarray}
&& R^{\dag}({\bf r}\tau)R({\acute{\bf r}}\tau)=\frac{1}{\sqrt{2}}\left[
\begin{array}{cc}
\sqrt{1+{\bf{\Omega}}(\textbf{r}\tau){\bf{\Omega}}(\textbf{r}'\tau)}
\exp(\frac{i}{2}\Phi)
 & \sqrt{1-{\bf{\Omega}}(\textbf{r}\tau){\bf{\Omega}}(\textbf{r}'\tau)}
\exp(\frac{i}{2}\bar{\Phi}) \\
-\sqrt{1-{\bf{\Omega}}(\textbf{r}\tau){\bf{\Omega}}(\textbf{r}'\tau)}
\exp(-\frac{i}{2}\bar{\Phi})  &\sqrt{1+{\bf{\Omega}}(\textbf{r}
\tau){\bf{\Omega}}(\textbf{r}'\tau)}\exp(-\frac{i}{2}\Phi)
\end{array}
\right],
\label{10}
\end{eqnarray}
where $\Phi\equiv\Phi[{\bf \Omega}({\bf r}\tau),{\bf \Omega}({\bf r'}\tau ),{\bf z}]$
is the signed solid angle spanned by the vectors
 ${\bf \Omega}({\bf r}\tau),{\bf \Omega}({\bf r'}\tau )$ 
and {\bf z} with $\bar{\Phi}=\Phi[{\bf \Omega}({\bf r}\tau),-{\bf \Omega}({\bf r'}\tau )]
-2\varphi({\bf r}\tau)$. In the complex projective representation, Eq.(\ref{10}) reads
\begin{eqnarray}
R^{\dag}({\bf r}\tau)R({\acute{\bf r}}\tau)
=\left[
\begin{array}{cc}
{\bar{\zeta}}_{1}(\textbf{r}\tau)\zeta_{1}(\textbf{r}'\tau)
+{\bar{\zeta}}_{2}(\textbf{r}\tau)\zeta_{2}(\textbf{r}'\tau)
 & -{\bar{\zeta}}_{1}(\textbf{r}\tau)\bar{\zeta}_{2}(\textbf{r}'\tau)
+{\bar{\zeta}}_{2}(\bf{r}\tau)\bar{\zeta}_{1}(\textbf{r}'\tau)\\
-{\zeta}_{2}(\textbf{r}\tau)\zeta_{1}(\textbf{r}'\tau)
+{\zeta}_{1}(\textbf{r}\tau)\zeta_{2}(\textbf{r}') &{\zeta}_{2}
(\textbf{r}\tau)\bar{\zeta}_{2}(\textbf{r}'\tau)+{\zeta}_{1}(\textbf{r}\tau)\bar{\zeta}_{1}(\textbf{r}'\tau)
\end{array}
\right].\ 
\label{R}
\end{eqnarray}
\end{widetext}
The form of Eq.(\ref{R}) suggests the use of the bond operators defined by Eqs.(\ref{aaa})
and (\ref{fff}), 
so that the product of rotational matrices can be written in a compact form
\begin{eqnarray}
&&R^{\dag}({\bf r}\tau)R({\acute{\bf r}}\tau)
=\left[
\begin{array}{cc}
{\cal F}
 & -\bar{\cal A}\\
{\cal A} &\bar{\cal F}
\end{array}
\right]({\bf{r}}\tau{\bf{r}}'\tau)
\nonumber\\
&&R^{\dag}({\acute{\bf r}}\tau{\bf r}\tau)R({\bf r}\tau)
=\left[
\begin{array}{cc}
\bar{\cal F}
 & \bar{\cal A}\\
-{\cal A} &{\cal F}
\end{array}
\right]({\bf{r}}\tau{\bf{r}}'\tau).
\end{eqnarray}
 With the help of the above equation it is easy to write down the components of the
$M$ matrix

Under the transformation the bond operators become
\begin{eqnarray}
{\cal{F}}({\bf{r}}\tau{\bf{r}}'\tau)&\to&{\cal{F}'}({\bf{r}}\tau{\bf{r}}'\tau)
\nonumber\\
&=&\frac{-\bar{\zeta}_{1}({\bf{r}}\tau)
{\zeta}_{2}({\bf{r}}'\tau)+\bar{\zeta}_{2}({\bf{r}}\tau){\zeta}_{1}({\bf{r}}'\tau)}{\sqrt{2}}
\nonumber\\
{\cal{A}}({\bf{r}}\tau{\bf{r}}'\tau)&\to&{\cal{A}'}({\bf{r}}\tau{\bf{r}}'\tau)
\nonumber\\
&=&
\frac{{\zeta}_{1}({\bf{r}}\tau){\zeta}_{}({\bf{r}}'\tau)
+{\zeta}_{2}({\bf{r}}\tau){\zeta}_{2}({\bf{r}}'\tau)}{\sqrt{2}}.
\end{eqnarray}
\begin{widetext}
\begin{eqnarray}
&&M_{12;21}({\bf{r}}
\tau,{\bf{r}}'\tau|{\bf{r}}'\tau,{\bf{r}}\tau)=\left\langle\left[R^{\dag}
({\bf{r}}\tau)R({\bf{r}}'\tau)\right]_{12}\left[R^{\dag}({\bf{r}}'\tau)
R({\bf{r}}\tau)\right]_{21}\right\rangle_{\text{SU(2})}
\nonumber\\
&&=\langle\bar{\cal{A}'}({\bf{r}}\tau{\bf{r}}'\tau)
{\cal{A}'}({\bf{r}}\tau{\bf{r}}'\tau)\rangle_{\text{SU(2)}}
=\left\langle\left[\bar{\zeta}_{1}({\bf{r}}'\tau)\bar{\zeta}_{1}({\bf{r}}\tau)
+\bar{\zeta}_{2}({\bf{r}}'\tau)\bar{\zeta}_{2}({\bf{r}}\tau)\right]
\left[{\zeta}_{1}({\bf{r}}\tau)\zeta_{1}({\bf{r}}')+\zeta_{2}
({\bf{r}}\tau)\zeta_{2}({\bf{r}}'\tau)\right]\right\rangle_{\text{SU(2)}}
\nonumber\\
&&=\left\langle\sum_{\alpha,\beta}\bar{\zeta}_{\alpha}({\bf{r}}'\tau)
\bar{\zeta}_{\alpha}({\bf{r}}\tau){\zeta}_{\beta}({\bf{r}}\tau)
{\zeta}_{\beta}({\bf{r}}'\tau)\right\rangle_{\text{SU(2)}}.
\end{eqnarray}
Furthermore, by implementing the Wick theorem to the $CP^1$ averages
\begin{eqnarray}
\left\langle \bar{\zeta}_{\alpha}({\bf{r}}'\tau)\bar{\zeta}_{\alpha}({\bf{r}}\tau)
{\zeta}_{\beta}({\bf{r}}\tau){\zeta}_{\beta}({\bf{r}}'\tau)\right\rangle_{\text{SU(2)}}&=&
\left\langle \bar{\zeta}_{\alpha}({\bf{r}}'\tau)\bar{\zeta}_{\alpha}
({\bf{r}}\tau)\right\rangle_{\text{SU(2)}}
\left\langle {\zeta}_{\beta}({\bf{r}}\tau)
{\zeta}_{\beta}({\bf{r}}'\tau)\right\rangle_{\text{SU(2)}}
\nonumber\\
&+&\left\langle\bar{\zeta}_{\alpha}({\bf{r}}'\tau)
\zeta_{\beta}({\bf{r}}\tau)\right\rangle_{\text{SU(2)}}
\left\langle\bar{\zeta}_{\alpha}({\bf{r}}\tau)
\zeta_{\beta}({\bf{r}}'\tau)\right\rangle_{\text{SU(2)}}
\nonumber\\
&+&\left\langle\bar{\zeta}_{\alpha}({\bf{r}}'\tau)\zeta_{\beta}({\bf{r}}'\tau)\right\rangle_{\text{SU(2)}}
\left\langle\bar{\zeta}_{\alpha}({\bf{r}}\tau)
\zeta_{\beta}({\bf{r}}\tau)\right\rangle_{\text{SU(2)}}
\nonumber\\
&=&\delta_{\alpha\alpha}\delta_{\beta\beta}
f(-{\bf d})f({\bf d})+\delta_{\alpha\beta}
\delta_{\beta\alpha}g({\bf d})g(-{\bf d})+
\delta_{\alpha\beta}\delta_{\beta\alpha}g({\bf 0})g({\bf 0})
\end{eqnarray}
\end{widetext}	
In a similar manner
\begin{eqnarray}
&&M_{11,11}({\bf{r}}
\tau,{\bf{r}}'\tau|{\bf{r}}'\tau,{\bf{r}}\tau)=2\left[g^2({\bf 0})-f^2({\bf d })\right],
\nonumber\\
&&M_{11;22}({\bf{r}}
\tau,{\bf{r}}'\tau|{\bf{r}}'\tau,{\bf{r}}\tau)=2\left[f^2({\bf 0})-g^2({\bf d})\right],
\nonumber\\
&&M_{12;21}({\bf{r}}
\tau,{\bf{r}}'\tau|{\bf{r}}'\tau,{\bf{r}}\tau)=4f^2({\bf d})+2g({\bf d})+2g({\bf 0}),
\nonumber\\
&&M_{12;12}({\bf{r}}
\tau,{\bf{r}}'\tau|{\bf{r}}'\tau,{\bf{r}}\tau)=-\left[6f^2({\bf d})+2f^2({\bf 0})\right],
\end{eqnarray}
where we have  used that $f(-{\bf d}) =f({\bf d})$ and $g(-{\bf d}) =g({\bf d}) $.
It is not difficult to see that
\begin{eqnarray}
M_{11,11}({\bf{r}}
\tau,{\bf{r}}'\tau|{\bf{r}}'\tau,{\bf{r}}\tau)=M_{22,22}({\bf{r}}
\tau,{\bf{r}}'\tau|{\bf{r}}'\tau,{\bf{r}}\tau),
\nonumber\\
M_{11;22}({\bf{r}}
\tau,{\bf{r}}'\tau|{\bf{r}}'\tau,{\bf{r}}\tau)=M_{22,11}({\bf{r}}
\tau,{\bf{r}}'\tau|{\bf{r}}'\tau,{\bf{r}}\tau),
\nonumber\\
M_{12;21}({\bf{r}}
\tau,{\bf{r}}'\tau|{\bf{r}}'\tau,{\bf{r}}\tau)=M_{21;12}({\bf{r}}
\tau,{\bf{r}}'\tau|{\bf{r}}'\tau,{\bf{r}}\tau),
\nonumber\\
M_{12;12}({\bf{r}}
\tau,{\bf{r}}'\tau|{\bf{r}}'\tau,{\bf{r}}\tau)=M_{21;21}({\bf{r}}
\tau,{\bf{r}}'\tau|{\bf{r}}'\tau,{\bf{r}}\tau),
\end{eqnarray}
while all the remaining components of the $M$ matrix  vanish.
%
\section{APPENDIX C:USEFUL OPERATOR IDENTITIES}
%
By introducing the fermionic representation of the $1/2$-spin operators
\begin{eqnarray}
{\bf{S}}_h({\bf{r}}\tau)=\frac{1}{2}
\sum_{\alpha\beta}\bar{h}_{\alpha}({\bf{r}}\tau)
\hat{\sigma}_{\alpha\beta}h_{\beta}({\bf{r}}\tau)
\end{eqnarray}
and the following bond operators:
\begin{eqnarray}
&&{\cal A}_h'({\bf{r}}\tau{\bf{r}}'\tau)=\frac{h_{\uparrow}({\bf{r}}\tau)
h_{\uparrow}({\bf{r}}'\tau)+h_{\downarrow}({\bf{r}}\tau)
h_{\downarrow}({\bf{r}}'\tau)}{\sqrt{2}},
\nonumber\\
&&{\cal A}_h({\bf{r}}\tau{\bf{r}}'\tau)=\frac{h_{\uparrow}({\bf{r}}\tau)
h_{\downarrow}({\bf{r}}'\tau)-h_{\downarrow}({\bf{r}}\tau)h_{\uparrow}({\bf{r}}'\tau)}{\sqrt{2}},
\nonumber\\
&&{\cal F}_h({\bf{r}}\tau{\bf{r}}'\tau)=\frac{\bar{h}_{\uparrow}({\bf{r}}\tau)
h_{\uparrow}({\bf{r}}'\tau)+\bar{h}_{\downarrow}({\bf{r}}\tau)h_{\downarrow}({\bf{r}}'\tau)}{\sqrt{2}},	
\end{eqnarray}	
one can prove the following  useful identities that involve four-fermion products
that appear in the second-order cumulant expansion. For the spin and charge products,
we have
\begin{eqnarray}
&&{\bf{S}}_h({\bf{r}}\tau)\cdot{\bf{S}}_h({\bf{r}}'\tau)=\frac{n_h({\bf{r}}'\tau)}{4}
\nonumber\\
&&-\frac{\bar{\cal A}_h({\bf{r}}\tau{\bf{r}}'\tau){\cal A}_h({\bf{r}}\tau{\bf{r}}'\tau)}{2}
-\frac{\bar{\cal F}_h({\bf{r}}\tau{\bf{r}}'\tau){\cal F}_h({\bf{r}}\tau{\bf{r}}'\tau)}{2},
\nonumber\\
&&\frac{n_h({\bf{r}}\tau)n_h({\bf{r}}'\tau)}{2}=\frac{n_h({\bf{r}}'\tau)}{2}
\nonumber\\
&&+\bar{\cal A}_h({\bf{r}}
\tau{\bf{r}}'\tau){\cal A}_h({\bf{r}}\tau{\bf{r}}'\tau)
-\bar{\cal F}_h({\bf{r}}\tau{\bf{r}}'\tau){\cal F}_h({\bf{r}}\tau{\bf{r}}'\tau).
\end{eqnarray}
Furthermore, for the prpducts of fermionic variables that appear in the calculation
of the $M$ matrix, one finds
\begin{eqnarray}
&&\sum_{\alpha\beta}\bar{h}_{\alpha}({\bf{r}}\tau)
h_{\alpha}({\bf{r}}'\tau)\bar{h}_{\beta}({\bf{r}}'\tau)
h_{\beta}({\bf{r}}\tau)=n_h({\bf{r}}\tau)
\nonumber\\
&&-\frac{n_h({\bf{r}}\tau)n_h({\bf{r}}'\tau)}{2}-2{\bf{S}}_h({\bf{r}}\tau)\cdot{\bf{S}}_h({\bf{r}}'\tau),
\nonumber\\
&&\sum_{\alpha\beta}\bar{h}_{\alpha}({\bf{r}}\tau)
h_{\beta}({\bf{r}}'\tau)\bar{h}_{\beta}({\bf{r}}'\tau)
h_{\alpha}({\bf{r}}\tau)=2n_h({\bf{r}}\tau)
\nonumber\\
&&-n_h({\bf{r}}\tau)n_h({\bf{r}}'\tau),
\nonumber\\
&&\sum_{\alpha\beta}\bar{h}_{\alpha}({\bf{r}}\tau)h_{\beta}({\bf{r}}'\tau)
\bar{h}_{\alpha}({\bf{r}}'\tau)h_{\beta}({\bf{r}}\tau)=n_h({\bf{r}}\tau)
\nonumber\\
&&-2\bar{\cal A}_h'({\bf{r}}\tau{\bf{r}}'\tau){\cal A}_h'({\bf{r}}\tau{\bf{r}}'\tau),
\end{eqnarray}
where, by simple inspection, one can find that  the spin-rotational symmetry is apparent.


\begin{references}
\bibitem{retard} 
It is evident  that the retardation plays a crucial role in the BCS mechanism.
In the typical metallic superconductor, the Fermi energy $E_F$ is of the order of several
$10$eV while phonon frequencies  $\omega_D$ are of the order $10^{-2}$eV, therefore
$E_F/\omega_D\sim 10^3$. Since
the renormalization of the Coulomb pseudopotential is logarithmic, this large
value of the retardation is required. In the cuprate superconductors, on the other hand,
$E_F/\omega_D <  E_F/2\Delta_0	\sim 5$, where $\Delta_0$ is the superconducting gap,
so the ``HTSC" materials are clearly in the nonretarded regime.
\end{references}
\end{document}